\documentclass[paper=A4,oneside]{article} 
\usepackage{amsmath}
\usepackage{amssymb}
\usepackage[utf8x]{inputenc}
\usepackage[T1]{fontenc}
\usepackage{graphicx}
\usepackage{color}
\usepackage{hyperref}
\usepackage{natbib}
\usepackage{authblk}

\begin{document}
\renewcommand{\baselinestretch}{1.0}\normalsize

\title{How John Wheeler lost his faith in the law}
\author[1]{Alexander Blum}
\author[2]{Stefano Furlan}
\affil[1]{Max Planck Institute for the History of Science, Berlin, and Albert Einstein Institute, Potsdam, Germany}
\affil[2]{Max Planck Institute for the History of Science, Berlin, and University of Geneva}
\maketitle

\begin{abstract}
In 1972, at a symposium celebrating the 70th birthday of Paul Dirac, John Wheeler proclaimed that „the framework falls down for everything that one has ever called a law of physics“. Responsible for this „breakage […] among the laws of physics“ was the general theory of relativity, more specifically its prediction of massive stars gravitationally collapsing to „black holes“, a term Wheeler himself had made popular some years earlier. In our paper, we investigate how Wheeler reached the conclusion that gravitational collapse calls into question the lawfulness of physics and how, subsequently, he tried to develop a new worldview, rethinking in his own way the lessons of quantum mechanics as well as drawing inspiration from other disciplines, not least biology.
\end{abstract}

The second half of the twentieth century saw the scientific discipline of physics at its prime. It also saw the beginnings of a perceived decline, with the life sciences being perceived more and more as the scientific avantgarde.\footnote{For some early remarks in this direction, see a letter from Wolfgang Pauli to Markus Fierz of 14 September 1954 \citep[p. 756]{meyenn_1999_wissenschaftlicher}. See also \citep{potthast_2007_was-bedeutet}.} One aspect of this development is the increasing doubt in the validity and relevance of reductionist, microscopic physical laws; another is the increasing importance in physical theorizing of concepts originating in the life sciences, such as emergence or evolution.\footnote{We will be talking about evolution in this chapter. For the story of how the notion of emergence entered the physical sciences, cf. \citep{munoz_2022_the-development}.} In this paper, we will study these developments by following the trajectory of a single, albeit prominent, scientist, the American physicist John Wheeler,\footnote{For biographical information on Wheeler, see his autobiography \citep{wheeler_1998_geons}.} who in the early 1970s began to promote the idea that the world was lawless in his foundations (culminating in his statement of ``law without law'' in the late 1970s) and to explore evolutionary ideas with respect to the cosmological genesis of physical laws.
 
Wheeler was hardly predestined to one day promote the fundamental lawlessness of physics. To the contrary. In the 1950s, Wheeler developed and promoted a methodology he called daring conservatism, which was based on taking the established laws of physics seriously, upholding them for as long as possible, and testing their implications to the utmost (that is the ``daring'' part). The hero of this approach was Niels Bohr, whose atomic theory Wheeler took to be the pinnacle of daring conservatism:

\begin{quote}
I know no better example of this daring conservatism than Bohr’s analysis of the hydrogen atom. Nothing was further from his approach than the free play with ideas that one finds in the Philosophical Magazine of the 1910’s. There one author treated atoms as electric fluids, another made a hydrodynamical model, others changed the laws of electrostatics, and still others altered the laws of electromagnetic radiation to account for the absence of radiation from stable atoms. In contrast, Bohr accepted the well founded law of force between point charges, made use of the usual principles of mechanics, and recognized that the laws of electromagnetic radiation are of far reaching application…\footnote{The quote is from a revised but unpublished version of Wheeler's 1954 Richtmyer Lecture to be found in the archive of Wheeler's papers at the American Philosophical Society Library (Mss. B.W564) in Philadelphia (which we will be referring to simply as ``Wheeler Papers'' in the following), Box 182, folder ``Fields and Particles''. See also \citep[footnote 40]{blum_2019_tokyo}.} 
\end{quote}

Some twenty years later, his views had changed, dramatically:

\begin{quote}
[T]he closed universe of Einstein’s general relativity […] undergoes gravitational collapse. In that collapse, classical space and time themselves come to an end. With their end, the framework falls down for everything that one has ever called a law of physics.\\
Nothing that relativity has ever predicted is more revolutionary than collapse, and nothing that collapse puts into question is more central than the very possibility of any enduring laws in physics. \citep[p. 203]{wheeler_1973_from-relativity}
\end{quote}

The story of this transformation, from arguing that the laws of physics should be extrapolated to the extreme to arguing that at bottom there are no fundamental laws, at first glance seems straightforward and in the second quote Wheeler is very explicit about the reasons for his newly-held beliefs: it is the realization that the universe will end in gravitational collapse, in what is now customarily referred to as a ``big crunch’’. Indeed, this seems to fit with popular notions, shaped primarily by big crunch’s little sibling, the gravitational collapse of a massive star into a black hole; after all, doesn’t everyone know that in black holes all bets are off?

But the story is not quite as simple. For starters, the popular imagination of black holes is to a large part shaped by the ideas of Wheeler himself (who famously promoted the use of the term ``black hole''). So one needs to be very careful when taking the radical implications of gravitational collapse as self-evident. In this chapter, we aim to outline how it was Wheeler’s rather idiosyncratic takes on general relativity, cosmology, and quantum theory that led him to draw such radical conclusions from gravitational collapse. In particular, we will reconstruct how Wheeler reached the conclusion (a) that \emph{all} known laws of physics were violated in gravitational collapse and (b) that \emph{no} newly discovered laws of physics would step in to replace them; both of these conclusions, it needs to be stressed again, are not at all straightforward.

We will be looking at Wheeler's changing views from several angles. In doing so, we will be covering the same historic period, from the 1950s to the 1970s, several times over. We thus offer here a very simple periodization of these decades. Wheeler's thinking is more dynamical than the sharp breaks implied by such a periodization might imply, all the more as he frequently pursued different and even \emph{prima facie} contradictory approaches in parallel; but some basic dates will still be helpful for the chronological orientation of the reader. In the period from 1952 to 1960 Wheeler first formulated his daring conservative program for general relativity and began investigating the question of gravitational collapse. He still expected that total collapse into what we would now call a black hole would not actually occur in nature. In the early 1960s, Wheeler came to accept that gravitational collapse might proceed past the so-called Schwarzschild horizon \citep{furlan_2021_john} and that a quantum theory of gravity would be necessary to provide an adequate description of gravitational collapse. This mature phase of his research program goes approximately from 1960 to 1969. Around 1970 he began to question the foundational assumptions of his research program and to embark on his exploration of lawlessness, the origins of which it is the purpose of this paper to explain.

We will begin in Section \ref{sec:baryon} by studying how Wheeler first reached the conclusion that in gravitational collapse established conservation laws of physics (most notably baryon number, which roughly corresponds to the amount of matter) would not just be violated, but, as he put it, ``transcended''. In Section \ref{sec:geometro}, we reconstruct why Wheeler rejected lawful alternatives for analyzing the process of gravitational collapse and how this led him -- in combination with the observed transcendence of conservation laws -- to promote the view that physical law ended in gravitational collapse. We continue in Section \ref{sec:cosmo} by showing how Wheeler's emerging views on the specific process of gravitational collapse affected his views on physics more generally by looking at his cosmology. In Section \ref{sec:bio}, we will give an outlook on Wheeler's ultimately unsuccessful attempts of the 1970s to harness concepts from the life sciences in order to flesh out what a lawless physics might look like, before we give some brief conclusions in Section \ref{sec:6}.

\section{Transcending Conservation Laws}
\label{sec:baryon}

John Wheeler's engagement with general relativity, and his identification of that theory as the ideal vehicle for his daring conservatism, begins in 1952 \citep{blum_2019_tokyo}. Early on, Wheeler identified the gravitational collapse of massive stars as one of the central physical problems he was interested in. After all, what better way to extrapolate Einstein's theory of gravity to its extremes than to apply it to the most dense (and thus most strongly gravitating) objects expected to exist in the universe. But of course general relativity was not the only physical theory involved in stellar collapse; after all stars were made of matter. And with regards to matter one law in particular appeared to Wheeler to be essential to the question of gravitational collapse: the law of baryon number conservation.

The idea that there was a conservation law for the amount of matter was of course an old one. Over the course of the first half of the 20th century, it had become clear that the bulk of the mass of (atomic) matter was concentrated in the nucleus, and that this nucleus was composed of protons and neutrons, collectively known as nucleons or heavy particles. In 1949, Eugene Wigner had recast the law of the conservation of matter as a ``conservation law for the number of heavy particles (protons and neutrons)'' \citep[fn. 9]{wigner_1949_invariance}. In the following years, other heavy particles were discovered in cosmic ray and accelerator experiments, and in 1953, Abraham Pais introduced the term ``baryon'' as ``a collective name for heavy particles which not only comprises nucleons'' \citep[p. 157]{pais_1954_omega}. The name caught on, and Wigner's law is now commonly referred to as ``baryon number conservation''. We shall also be using this term, even though the baryons Wheeler was dealing with were only nucleons and he himself referred to them only as nucleons or heavy particles.

Wigner had introduced the law of baryon number conservation as a conjecture to explain the fact that the proton (and thus atomic matter in general) does not decay radioactively. Within a few years, the law of baryon number conservation had become a mainstay of particle physics \citep{borrelli_2015_the-making}. But what, precisely, was Wheeler's interest in this law?

A central issue in the physics of stellar collapse was which state a star would settle into in after having burned out its fuel and collapsing under its own weight. The 1930s had thus seen numerous studies of the equilibrium configurations of superdense, cold matter \citep{bonolis_2017_stellar}. The densest configuration of matter that had been studied was a star (or a stellar core) consisting only of neutrons. Taking into account general-relativistic effects in the star's self-gravitation, J.R. Oppenheimer and George Volkoff had found a limiting mass of 0.71 solar masses, above which there was no equilibrium state for a collection of neutrons \citep{oppenheimer_1939_on-massive}. 

Oppenheimer and Volkoff had concluded that stars with a greater mass would not be able to reach  hydrostatic equilibrium and would collapse indefinitely in a process further studied by Oppenheimer and Hartland Snyder, a paper now considered a pioneering work on the formation of black holes \citep{oppenheimer_1939_on-continued}. But Wheeler was skeptical of Oppenheimer’s conclusion, because it had not taken into account the conservation of baryon number. Oppenheimer and Volkoff had not talked much about baryon number conservation, for the simple reason that it really hadn’t been a thing yet in 1939. But Wheeler saw that taking into account the law of baryon number conservation would have striking consequences for neutron stars in particular and stellar collapse in general. That realization was made possible by the conceptual changes that physics had undergone in the intervening 15 years; but it also reflects an ambition that, in different forms, guided Wheeler's research throughout decades: that of tying together the extremely large and the extremely small. This aspiration found highly speculative expressions already in his early notebooks („Most exciting idea of each particle being a universe, perhaps identical with our own“; "Looks like a proton on the outside, but is a whole universe on the inside“).\footnote{These quotes are from Wheeler's first Relativity notebook, p. 128. We will quoting frequently from Wheeler's series of Relativity notebooks, which contains 19 volumes spanning from 1952 to 1973 and is conserved at the Wheeler Papers. We will be referring to the individual notebooks as NBX, where X is the number of the notebook.} But it could also manifest itself far more concretely (and we shall see more instances of this further on), as here in the attempt of bringing together the microscopic law of baryon number conservation with the macroscopic process of stellar evolution.

Of course, the gap between micro- and macrophysics was not easily bridged: the conservation of baryon number was observed empirically only for the small nuclei available to terrestrial experimentation, not for the gigantic agglomerations of neutrons that might be found in stellar cores. But bridging such gaps was a central part of Wheeler's program of daring conservatism. So, when Wheeler re-analyzed the work of Oppenheimer and Volkoff  by ``simply recognizing the `law’ of heavy particle conservation’’ (1 April 1956, NB4, p. 36), he was conceding (with his quotation marks) that the `law’ of baryon number conservation might not be as time-honored as Coulomb’s law, but that its extrapolation to the stellar scale could still be considered an instance of daring conservatism.

What were now the implications of considering the law of baryon number conservation for neutron cores? It offered an alternate reading of the Oppenheimer-Volkoff limit. Oppenheimer had assumed that the neutron core would become unstable if mass was added beyond the limit. This was illustrated by looking at the free energy as a function of density for different values of the mass: equilibrium was attained for the density minimizing the free energy, but for masses greater than the limiting value the free energy curve did not have any minima. Wheeler, in his first reflections on the Oppenheimer-Volkoff paper, objected to this illustration:

\begin{quote}
Discusses free energy but incorrectly in my opinion since figure free energy as a function of $t_0$ [effectively the density at the center of the neutron core] not for fixed no. of particles but for fixed mass. (NB4, p.35)
\end{quote}

Wheeler was thus arguing (to himself in his notebook) that mass was a bad measure of the fixed amount of matter contained in a collapsing star. After all, mass was already in special relativity something that could be transformed into energy and thus radiated away. A better gauge of the amount of matter was baryon number. But if one consequently imagined making a star bigger not by increasing its mass, but by increasing the number of baryons there was a whole different way to read the Oppenheimer-Volkoff limit: not as a maximal mass for stability, but as a maximal mass, period. What this implied was that if one had reached the maximal mass and then added another nucleon, the mass would no longer increase; instead it would presumably decrease (the mass of the nucleon plus the mass lost by the star would be radiated away), ultimately reaching some constant mass value. 

\begin{figure}
  \includegraphics[width=\linewidth]{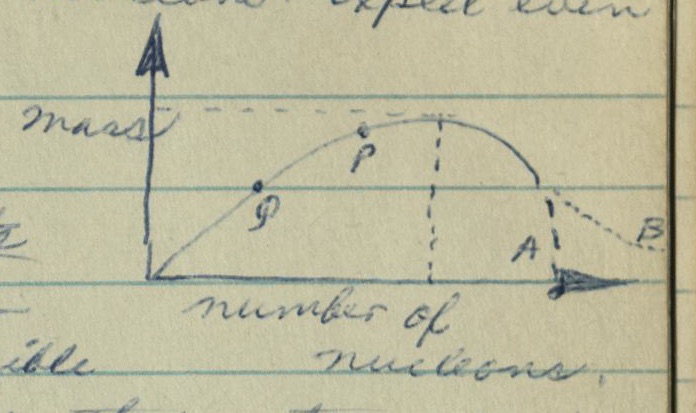}
  \caption{Mass of star as function of the number of nucleons. The maximal mass is the critical mass of Oppenheimer and Volkoff. Option A represents the possibility that the star reaches a zero mass value already for a finite number of nucleons. In option B zero mass is achieved only asymptotically, for an infinite number of nucleons. NB4, p. 39, 1 April 1956.}
  \label{fig:nb4}
\end{figure}

The most striking case, which Wheeler briefly appears to have liked best, would be if that constant mass were zero and if the star would reach that mass already for a finite number of baryons (NB4, p. 39, see figure \ref{fig:nb4}). Adding further nucleons to such an object would make no difference to the mass (``all the mass energy will be radiated’’) and no appreciable difference at all, except to change the baryon number. And while this would not change any of the immediate properties of the object itself, it might eventually make ``a difference’’: Wheeler imagined ``a special one of the gods who will keep track of the number of nucleons in the thing - no one else will care.’’ When the ``limiting object“ then encountered a limiting ``anti-object’’ (with a large, supercritical negative baryon number), the two would annihilate and result in an object with a possibly subcritical and thus observable baryon number: if, e.g., the the number of baryons in the first object minus the number of anti-baryons in the second were 1, the result of their annihilation would be a neutron or a hydrogen atom with baryon number 1. Wheeler imagined that ``the special god would smile and say `Look, see!’’’ — vindicated, one might say, in his daring-conservative belief in the law of baryon number conservation.

But could such super-condensed matter really exist? While Wheeler briefly entertained the possibility, he had discovered a paradoxical thought experiment, which ultimately appears to have turned him away from supercondensation:

\begin{quote}
When one has nearly the peak mass for stability against collapse, one has a very interesting situation — add one nucleon; mass of star increases by a negligible fraction of mass of one nucleon; hence bulk of mass is given off as heat. When beyond this point, get off more energy than $MC^2$; hence could have 2 such stars, one the anti of the other and make a profit producing nucleons out of vac[uum]. Expect star will take profit. (NB4, p. 36)\end{quote}

In other words, one could create a nucleon-antinucleon pair from the vacuum, send the nucleon into a critical star and the anti-nucleon into a critical anti-star and get back more energy than was needed to create the pair.\footnote{It should be noted that this is not a perpetual motion machine, as the two stars (star and anti-star) only have finite mass-energy to give away. After they have settled into a constant (possibly zero) mass state, Wheeler's process would only increase the baryon number of the star and the negative baryon number of the anti-star, with no energy being required or gained. Wheeler still considered this process an ``idealized machine to manufacture matter at no cost in energy'' \citep[p. 129]{adams_1958_some-implications} and an argument against baryon number conservation.}

Wheeler had worked out these ideas on a trip from Leiden,\footnote{On Wheeler's stay in Leiden, see \citep{heuvel_2020_together}.} where he was spending the year as Lorentz professor, to the US, where he was attending the Rochester conference. On the penultimate leg of his journey, on 2 April 1956, from Washington to his home in Princeton, he reached the conclusion that this was a rather fanciful conjecture and stated (NB4, p. 40):

\begin{quote}
``\emph{Either} there exists a super condensed phase \emph{or} law of conservation of nucleons fails — and we don’t see former’’ — This would be my first go around at seeking a conclusion.
\end{quote}

Wheeler devoted considerable energy and manpower to stellar collapse over the two following years, but this conclusion remained: no matter the exact dynamical details of gravitational collapse it would involve a violation of the law of baryon number conservation, a point he further emphasized in his talk at the 1958 Solvay conference \citep{adams_1958_some-implications}. The question of how exactly this violation would occur occupied him for many years. Initially, Wheeler had assumed that the process that violated baryon number would simultaneously prevent the collapsing star from shrinking below its Schwarzschild radius. This meant that the process of baryon number violation would cause the entire star to dissociate, which was the only way to prevent further collapse. Wheeler envisioned that the neutrons in the neutron star might turn into massless neutrinos (NB5, p. 135). This would only happen under the high densities obtained in gravitational collapse and would ``uphold[...] most other conservation laws'', such as the conservation of charge and angular momentum \citep[p. 127]{wheeler_1961_geometrodynamics}.

In the early 1960s, Wheeler came to accept the possibility that astronomical objects might collapse beyond their Schwarzschild radius. This did not, however, remove the issue of baryon number; indeed, the Schwarzschild radius had played no role in Wheeler's argument for the violation of baryon number conservation. But now Wheeler no longer had to assume that the whole star exploded; rather, he now envisioned that in collapse the nucleons were ``crushed out of existence'', losing their integrity as matter through a new form of radioactive decay and the emission of ``$\delta$-radiation'' (named in analogy to the three known forms of radioactivity, $\alpha$, $\beta$, and $\gamma$ radiation).\footnote{Notes taken by Dieter Brill of a talk given by Wheeler at the Texas Symposium on Relativistic Astrophysics, 16-18 December 1963. Personal possession of Dieter Brill. We would like to thank Dieter Brill for making these notes available to us.} Wheeler appears to have envisioned that the $\delta$ radiation actually carried away the baryon number, thereby ultimately implying a conservation of that quantity. It remained unclear, however, what the $\delta$ rays would do with that baryon number, i.e., how they would interact with other matter, so that this detail was ultimately inconsequential. And, in any case, these speculations were short-lived. Kip Thorne recalls that he ``discouraged'' Wheeler from including them in a 1965 book they co-authored \citep[p. 15]{thorne_2019_john}, which then only contained somewhat vague references to a new form of radioactivity in the final pages \citep{harrison_1965_gravitation}.

Specific suggestions aside, we see that Wheeler was looking for a \emph{mechanism} with which baryon number conservation was violated in gravitational collapse. He was thus looking for a new law to take the place of baryon number conservation. In the late 1960s, however, he came to realize that the violation of baryon number conservation in gravitational collapse might occur in a very different manner. The mid-1960s were the beginning of what Kip Thorne has called a ``golden age of black hole research''\footnote{\citep[p. 258]{thorne_1994_black}. Thorne credits William H. Press for the expression.} and saw a flurry of new theoretical results. We will mention the so-called singularity theorems later on. But more important for Wheeler's attitude toward baryon number conservation was the gradual emergence of the so-called (by Wheeler) no-hair theorem, that a star when collapsing to a black hole will lose almost all of its particularities\footnote{Indeed, before Wheeler settled on the name no-hair, he would speak of no-particularities, see \citep{wheeler_1969_from-mendeleevs}, where the image on p. 190 already contains the new phrasing ``A black hole has no hair'', while the text (which was presumably strongly based on the talk given two years earlier) has, on p. 191, ``A black hole has no particularities''.} and be characterized only by its mass (which determines its size), its total charge, and its angular momentum. All other characteristics will be washed out in collapse, such as complicated charge distributions that might lead to dipole moments or magnetic fields.

The first steps toward the theorem had been made by Werner Israel in early 1967, who had shown that a non-rotating, non-charged star would collapse into a perfectly spherical black hole, characterized only by its mass.\footnote{\citep{israel_1967_event}. See also \citep[p. 279]{thorne_1994_black} for more on the history of the no-hair theorem.} The completion of the theorem to include charge and rotation (and exclude anything else) was still a few years off, but Wheeler conjectured early on that this would eventually be proven (though, as we shall see, he still thought that magnetic moment might be included in the list of properties of black holes). And this raised an immediate question: what about baryon number? The question was all the more pressing as particle physicists had, in the interceding years, established further conservation laws of similar type, in particular the conservation of lepton number, the number of \emph{light} particles, such as electrons and neutrinos (which Wheeler consequently also referred to as neutrinoicity). Wheeler addressed this question in his notebooks (NB15, p. 60, 7 July 1968):

\begin{quote}
 The Great Question\\
 A few days ago trying to formulate my deepest concerns [...], I said, the great question is, do black holes and elementary particles belong to the same breed of cats. A black hole has mass, angular momentum, charge, magnetic moment -- but so far as I know, nothing else. Baryon number, neutrinoicity, strangeness -- all these as far as I can see, go out in the collapse.
\end{quote}

He then re-iterated the thought experiment with the special god, though now using neutrinos instead of baryons (NB 15, p. 61):

\begin{quote}
If a black hole ``made of'' neutrinos collides with one ``made of'' antineutrinos, is the outcome different from the collision of the black hole ``made of'' neutrinos and one of the same macroscopic quantities made of [neutrinos], or photons, or ordinary matter, or some mixture thereof?
\end{quote}

He was beginning to suspect that even if matter went into the black hole unharmed (i.e., without losing its identity through something like delta radiation, or as he put it in his notebook ``all down, none out''), the special god would no longer be vindicated in keeping track of how many baryons or neutrinos went in; baryon number as a property of matter would simply vanish in collapse since it was not a possible property of black holes, no special mechanism needed: ``all junk dropped down a black hole leaves it smooth'' (NB15, p. 106, 11 August 1968).\footnote{A little further on, in December 1968, we also find the first use of the expression ``no-hair'' (NB15, p. 213), though Wheeler would not start promoting the name until somewhat later.}

The way in which baryon number conservation was violated in black holes thus appeared to be very different from what Wheeler had expected, and indeed from earlier cases of putative laws of nature being violated in new phenomena and by new effects. One year later, in a talk given in Italy on the occasion of the centennial of Mendeleev's periodic table, Wheeler's opinion had solidified \citep[p. 192-194]{wheeler_1969_from-mendeleevs}:

\begin{quote}
Gravitational collapse deprives baryon number and lepton number of all significance. ``Baryon number is conserved.'' ``Lepton number is conserved.'' Of all principles of physics these familiar conservation laws belong among the most firmly established. Yet with gravitational collapse the content of these conservation laws also collapses. The established is disestablished.
\end{quote}

But he wanted to be entirely sure and devoted detailed studies to both the conservation of lepton number \citep{wheeler_1971_transcending} and of baryon number, the latter conducted by his PhD student Jacob \citet{bekenstein_1972_baryon}. Based on these studies, Wheeler arrived at a twofold argument. One, there was no possibility of seeing the particle fall into the black hole (or preventing it from falling in), because even light was not fast enough to catch up with a particle on its way in. And two, there was no way of determining the lepton or baryon number of a black hole after the fact: while the charge of a black hole could be determined in principle by the electrostatic field at large distances from the black hole, the quanta mediating the shorter-range (weak and strong nuclear) interactions that characterized leptons and baryons beyond their electric charge would not be able to cross the black hole's horizon:

\begin{quote}
It is true that no attempt to observe a change in baryon number has ever succeeded. Nor has anyone ever been able to give a convincing reason to expect a direct and spontaneous violation of the principle of conservation of baryon number. In gravitational collapse, however, that principle is not directly violated; it is transcended. It is transcended because in collapse one loses the possibility of measuring baryon number, and therefore this quantity can not be well defined for a collapsed object. \citep[p. 36]{ruffini_1971_introducing}
\end{quote}

In gravitational collapse, something happened to the conservation laws of particle physics that could no longer be called violation and for which Wheeler coined the term ``transcendence''. The validity of these laws could no longer be ascertained \emph{in principle}. We can clearly see this putting a dent in the applicability of physical laws in gravitational collapse. It is thus an essential starting point for Wheeler's embrace of lawlessness, and he would later invoke it as such \citep{wheeler_1973_from-relativity}. But of course the question of particle-physics conservation laws is somewhat peripheral to the question of gravitational collapse, which as a physical process is primarily governed by general relativity, the physical theory that Wheeler had placed at the center of his daring conservatism. We will thus discuss in the next section how general relativity itself got caught up in the lawlessness of gravitational collapse.

\section{The Rise and Fall of Geometrodynamics revisited}
\label{sec:geometro}

In the mid-1950s, Wheeler embarked on his daring-conservative, general-relativity-based research program, which he referred to as ``geometrodynamics''.\footnote{On Wheeler's program, as described in this section, see \citep{wheeler_1964_geometrodynamics}. Wheeler would sometimes equivocate and use ``geometrodynamics'' to refer to (a specific formulation of) general relativity, rather than to his research program of explaining the nature of elementary particles from geometry. \citet{misner_1974_some-topics} would differentiate between ``visionary geometrodynamics'' (Wheeler's program) and a more mundane ``physical geometrodynamics'' (along with a more formal ``mathematical geometrodynamics''). We will use the term ``geometrodynamics'' to refer to ``visionary geometrodynamics'' throughout.} From the start, this entailed more than just applying general relativity to extreme physical situations, such as gravitational collapse. Wheeler's hope had always been to use general relativity to construct a new foundation for microscopic particle physics. Particulate matter was to be reconceptualized as consisting merely of gravitational and electromagnetic fields concentrated within a small region. These particle-like field disturbances went by the name of ``geon'' (gravitational-electromagnetic entity) and their localized field energy could be interpreted as the mass of a particle; ``mass without mass'' in Wheeler's words. Similarly, electric charge was to be reconceptualized in terms of the topology of space; there would be handles connected to the bulk of space at two points and electric field lines trapped in those handles would emanate from those points as if there were an electric charge present. These handles were called ``wormholes'' and possessed ``charge without charge''.

It was clear on many levels that these structures built from classical fields were fully inadequate to represent (let alone explain the existence of) the known elementary particles. One might mention the fact that Wheeler was able to derive an upper limit for the charge-to-mass ratio of a wormhole on the order of $10^{-10} C/kg$ \citep[p. 593]{misner_1957_classical}; the electron has a charge-to-mass ratio on the order of $10^{11} C/kg$. Another issue was the instability of these structures. Indeed, numerical calculations performed by Wheelers's former PhD student Richard Lindquist together with Susan Hahn of IBM indicated that the ``throat'' of a wormhole would inevitably collapse and leave behind a singular configuration with infinite spacetime curvature \citep{hahn_1964_the-two}. Geons originally appeared to gradually vanish by leaking their energy \citep[p. 144]{wheeler_1961_geometrodynamics} until they were believed to collapse as violently as massive stars.\footnote{See NB15, p. 178, 10 November 1968: ``Do I not now believe that geons are probably \emph{unstable} with respect to \emph{collective} gravitational collapse...''}

Wheeler was thus convinced early on that a \emph{quantum theory} of geometrodynamics would be needed in order to turn his ideas for ``mass without mass'' and ``charge without charge'' into viable (and stable) models of elementary particles. This expectation was further reenforced by the so-called singularity theorem, established by Roger \citet{penrose_1965_gravitational}, which stated that in the gravitational collapse of stars (and of the universe, as proven by Stephen \citet{hawking_1966_the-occurrence}) singular points with infinite space-time curvature would occur. Wheeler anticipated that quantum theory would be able to prevent these instabilities and singularities, just as it had prevented the collapse of the hydrogen atom through Bohr's concept of the stationary state.

This expectation, that a quantum theory of gravity will be required to deal with the apparently singular behavior of gravitational collapse, is still widely held. Under this assumption, gravitational collapse would be very similar to blackbody radiation: a phenomenon where the current theory (general relativity or electrodynamics) gives nonsensical results (singularity or ultraviolet catastrophe) and this situation is alleviated by introducing a new (quantum) theory to replace it.\footnote{As is well known, this is \emph{not} the order of events in the historical discovery of quantum theory.} There would be nothing very interesting happening here, as regards the status of laws of nature. At best one could ponder the status of approximate laws and the relation between the old and the new theories.


Why then did Wheeler ultimately invoke gravitational collapse as the reason for giving up the (quantum) geometrodynamical program and embracing lawlessness? Wheeler's abandonment of geometrodynamics was first analyzed -- essentially while it was happening -- by John \citet{stachel_1974_the-rise} in a paper fittingly entitled ``The Rise and Fall of Geometrodynamics''.\footnote{A story similar to Stachel's, but in more compact form, is given by Aaron \citet[p. 153]{wright_2014_more-than-nothing}, who describes Wheeler as being ``dissatisfied with progress'' in his program.} Here, Stachel listed a number of ``conceptual difficulties'' (p. 39) that led Wheeler to abandon his program in the early 1970s. The central difficulty he highlighted was that of ``spin without spin''. In order to craft particles from fields, Wheeler not only had to reproduce their mass and their charge, but also their spin. Wheeler realized that (a) spin could be constructed from geometry if dynamical changes in the topology of space were possible, but that (b) such topological changes were disallowed by ``classical differential geometry'', forcing him to abandon all hope that geometrodynamics would be able to provide realistic particles with half-integer spin. 

There is certainly some truth in this story: Wheeler's ambitions could not be realized within the restricted framework of general relativity, so he gave up on the program of laying a new geometrical foundation for the constitution of matter. But there are some confusing elements in this story: Wheeler had pointed out the difficulty of ``spin without spin'' already in the early 1960s \citep{wheeler_1963_quantum}; what had changed 10 years later? Note further that this story then has nothing at all to do with gravitational collapse. And it gives no indication as to why Wheeler would go on to propagate, as quoted in the introduction, the breakdown of ``the very possibility of enduring laws in physics.'' We will thus in the following attempt to show that, to understand Wheeler's turn to lawlessness, one must understand the ``fall of geometrodynamics'' not as Wheeler abandoning his program because it could not be realized within the restricting framework of quantum gravity, but as Wheeler abandoning quantum gravity because it could not accommodate his program.

This may seem somewhat strange at first. How can one ``abandon'' quantum gravity, the apparently necessary reconciliation of general relativity and quantum mechanics? To understand this, we need to take a step back again to look more closely at Wheeler's program of quantum geometrodynamics and its wider context. For Wheeler was not the only one pursuing the question of quantum gravity; there was in fact a diversity of approaches to this subject, already in the first decades after World War II. We can ignore for our story the approach then prevalent among particle physicists, of transforming general relativity into a quantum theory of ``gravitons'' in full analogy to the photons of quantum electrodynamics. Even within the emerging relativity community there were (at least) two starkly different attitudes toward quantum gravity, which we can associate with two physicists, Wheeler and Bryce DeWitt \citep{blum_2017_the-1957}. 

For DeWitt, quantum gravity was a problem posed by the simultaneous presence in physical theory of the ``two tree-like giants, the quantum theory, and the general theory of relativity'' \citep[p. 64]{dewitt-morette_2011_the-pursuit}. He made ``no apology'' for pursuing this problem for in his ``opinion it is sufficient that the problem is there, like the alpinist's mountain.'' \citep[p. 377]{dewitt_1957_dynamical}. There were substantial formal and conceptual difficulties involved in combining quantum theory and general relativity, these had to be conquered, and then one would see what the result would be. Wheeler also appreciated the grandeur of reconciling the two great theories of modern physics and would later speak of them as "the two skyscrapers that overpower every other structure in the city.“\footnote{The quote is from ``The chemistry of geometry'', p. 847. This paper is from 1969 or 1970 and was apparently intended for use as a chapter in \citep{misner_1973_gravitation}. It can be found at \url{https://jawarchive.files.wordpress.com/2012/02/thechemistryofgeometry1969.pdf}.} But for Wheeler, this reconciliation was not an end in itself; rather it was an opportunity to further develop the guiding idea of geometrodynamics and thus, as we have outlined, to give a new foundation for all of physics.

More specifically, quantum gravity provided the means to stabilize his geometrodynamically constructed particles and ensure that they were not far too big. It also provided plausibility to his ideas of constructing particles through the topological features of space. He believed that a quantum theory of gravity would involve microscopic fluctuations not just in the metric structure of space-time, but also in its topological structure, resulting in spacetime acquiring a ``foam-like structure'' at a microscopic level \citep[p. 609]{wheeler_1957_on-the-nature}.\footnote{Wheeler did not initially use the term ``quantum foam'' for this foam-like structure, but ultimately adopted it after it had entered common usage and employed it in the title of his autobiography \citep{wheeler_1998_geons}.}\footnote{It should be mentioned that the length scale of quantum gravity, what Wheeler would later christen the Planck scale, was actually too \emph{small} to be associated with the known elementary particles. Wheeler was thus forced to imagine those elementary particles to be ``a collective state of excitation of the foam-like medium''. He was confident that the large gap between the scale of the known particles and the Planck scale would arise in analogy to superconductors, where the transition to superconductivity is coherent on scales of $10^{-4}$ cm \citep{pippard_1953_the-coherence} much larger than the lattice spacing of $10^{-8}$ cm. We would like to thank Rocco Gaudenzi for helping us to clarify this point.} One would thus have microscopic virtual wormholes popping in and out of existence, just like the virtual particles of high-energy physics. While this did not prove that charged particles were related to wormholes it did imply, as he remarked in his notebook, that wormholes were ``not an opium dream'' (NB4, p. 15, 14 March 1956). Wheeler's speculations also went beyond the realm of particle physics: he believed, for example, that quantum gravity would imply fluctuations in the dimensionality of space and might eventually explain why macroscopic spacetime has four dimensions, no more, no less \citep{blum_2022_the-interpretation}.

For a while it looked as if these two visions would be referring to the same theory, just to different approaches. DeWitt (and others, such as Peter Bergmann) was trying to reach a quantum theory of gravity by applying the quasi-algorithmic procedure of quantization to Einstein's theory of gravity, ironing out the many formal difficulties encountered on the way. Wheeler was cutting to the chase, assuming central features of quantum gravity from the outset (such as quantum foam), and pondering how they might be used to lay a new foundation for all of physical theory. While convergence of these approaches was not out of the question, the differences in the two approaches were quite manifest. Witness the exchange between Wheeler and DeWitt after the former's 1963 talk on gravitational collapse and Delta Rays, mentioned in the last section:

\begin{quote}
Bryce: Why not quantize imploding object instead? I can get Friedman universe to bounce! (after down to $10^{-33}$ cm)\\
Wheeler: [I] couldn't agree more, but don't we expect elementary particle phenomena forced upon us [?]\\
Bryce: But it doesn't require anything new!
\end{quote}

DeWitt was satisfied with quantum gravity filling the apparent gap in physical theory encountered in the central singularity of gravitational collapse, by predicting a rebound and avoiding the appearance of infinite quantities. For Wheeler, on the other hand, merely preventing the appearance of a singularity was not sufficient, though he was certain that quantum foam would take care of the issue \citep[fig. 3]{fuller_1962_causality}. Wheeler expected from the study of gravitational collapse and quantum gravity deep insights into the structure and physics of elementary particles \citep[p. 271]{wheeler_1968_superspace}, another instance of Wheeler's belief in a connection between the macroscopic and the microscopic.

In the late 1960s, however, the program of quantizing gravity reached an unexpected degree of concreteness, or, as Wheeler put it, ``maturity'' (NB14, p. 19, 30 April 1966), as DeWitt constructed a Schr\"{o}dinger-type equation for quantum gravity \citep{dewitt_1967_quantumI}, an equation that would become known, quite ironically, as the Wheeler-DeWitt equation. This was of course hardly a complete theory of quantum gravity, as many essential mathematical questions regarding this equation were (and are) still open. Even so, it clearly revealed that Wheeler's expectations would not be met in this approach. The Wheeler-DeWitt equation described only changes in the metric structure of space, not in its topology; and the dimensionality of spacetime, rather than appearing as a dynamical quantity, appeared to be hard-wired into the equation. Of course this was to be expected from a theory that arises from a direct quantization of general relativity; Wheeler liked to emphasize that the Wheeler-DeWitt equation can be derived directly from the Hamilton-Jacobi formulation of general relativity,\footnote{See, e.g., the preparatory notes for a talk in Munich, dated 14 July 1966, in NB14, p. 54: ``Not very diff[erent] in gen[era]l character from way do other q[uantum] prob[lem]s. Outline Hamilton Jacobi. Give DeWitt eq[uation].'' The Hamilton-Jacobi formulation of general relativity was developed by Asher \citep{peres_1962_on-cauchys}.} just like the Schr\"{o}dinger equation can be derived directly from the Hamilton-Jacobi formulation of classical mechanics.\footnote{It should be noted that this is not the derivation used by DeWitt in the paper introducing the Wheeler-DeWitt equation.} But the absence of a dynamical topology (and dimensionality) flew in the face of Wheeler's expectations for quantum geometrodynamics.

At first Wheeler was undeterred, and we find him describing his ``pursuit of gen rel, how I still had great hopes'' (NB14, p. 15, 9 March 1966), and more specifically imagining ways how to get topology change back in through the integration of electrodynamics:

\begin{quote}
Could it be that electromagnetism has to do with slipover from one topology to another; that that [sic] -- at quantum level -- ``puts ticks enough into state function''... (NB14, p. 20, 30 April 1966)
\end{quote}

But Wheeler struggled to establish a connection between the emerging formalism and his program. He articulated the frustration he felt at this gap when preparing a talk for the end of his presidency of the American Physical Society (NB14, p. 114, 4 January 1967):

\begin{quote}
The general story of the dynamics I know insofar as Hamilton Jacobi formalism is involved. And why isn't it enough to tell that. One isn't under obligation to answer all possible questions! [...] Many can ask, do humans have right to wade in on such ``high and holy issues''. But what alternative do we have? \\
Are obliged to discuss the relation to elementary particles? Einstein's dream?
\end{quote}

He soldiered on for a while. Topology change was central to his entire research program, providing a potentially unifying picture of macroscopic gravitational collapse and microscopic quantum foam:

\begin{quote}
Collapse is going on all the time at the virtual level. Every change of topology has in some sense a connection with collapse. (NB14, p. 120, 8 January 1967)
\end{quote} 

Wheeler thus tried somehow to fit changes in topology into the new framework:

\begin{quote}
Make a comment about topology changes analogous to that which Eddington made about temp[erature] required for thermonuc[lear] reations when he said: if center of sun isn't hot enough, go and find a hotter place. \emph{Must} take place. Kick the math to fit the phys[ics], not conversely. (NB14, p. 141, 27 January 1967)\footnote{See also NB15, pp. 33ff, 20 May 1968.}
\end{quote}

He also tried to formulate the Wheeler-DeWitt equation without reference to the dimensionality of space (NB15, p. 106ff, 11 August 1968). But ultimately it became clear all this would not work out; the new framework was too close to classical general relativity, which knew nothing of changing topologies and dimensionalities. Wheeler gradually had to give up the hope that quantum gravity could provide the framework for his vision of quantum geometrodynamics. In this sense, Stachel's story of the fall of geometrodynamics is correct. But this does not mean that he accepted the limits of the newly emerging formalism and tried to work within them. What he did instead is best described in Wheeler's own words by an allegorical story he included in the 1973 textbook \emph{Gravitation}, which in many ways represents the sum of his two decades working on general relativity:

 \begin{quote}
 There are reputed to be two kinds of lawyers. One tells the client what not to do. The other listens to what the client has to do and tells him how to do it. From the first lawyer, classical differential geometry, the client goes away disappointed, still searching for a natural way to describe quantum fluctuations in the connectivity of space. Only in this way can he hope to describe electric charge as lines of electric force trapped in the topology of space. Only in this way does he expect to be able to understand and analyze the final stages of gravitational collapse. Pondering his problems, he comes to the office of a second lawyer with the name ``Pregeometry'' on the door. Full of hope, he knocks and enters. What is pregeometry to be and say? Born of a combination of hope and need, of philosophy and physics, and mathematics and logic, pregeometry will tell a story unfinished at this writing, but full of incidents of evolution so far as it goes. \citep[p. 1203]{misner_1973_gravitation}
 \end{quote}
 
The reference to ``classical differential geometry'' is somewhat striking. In 1967, Wheeler had already been well aware of these limits \citep{geroch_1967_topology},\footnote{See also NB14. p. 120-121, 8 January 1967, which contains notes of a phone call with Geroch concerning the impossibility of topology change.} but could still anticipate that ``[a] new world opens out for analysis in quantum geometrodynamics.'' \citep[p. 284]{wheeler_1968_superspace}; a few years later, it was enough to invoke the limits of classical differential geometry, for the limits of a quantum theory of gravity based on the Wheeler-DeWitt equation were just the same as those of the classical theory. \footnote{This view of quantum gravity as hopelessly rooted in the classical theory is made most explicit in \citep[Section 4]{patton_1975_is-physics}. Even here we see Wheeler using scare quotes when employing the term ``quantum gravity'' for a theory that ``take[s] as given Einstein's Riemannian space and his standard classical geometrodynamics law''.} We also see that Wheeler -- faced with the limits of classical differential geometry and a quantum theory of gravity that was not able to overcome these limits -- did \emph{not} give up on his program, but instead turned to a different ``lawyer'' who would tell him ``how to do it.''

We will talk more about the new lawyer's suggestions in the following sections, but we first take a moment to assess how all of this ties in with gravitational collapse and the embrace of lawlessness. In the previous section, we have seen that Wheeler had established a connection between gravitational collapse and the transcendence of microscopic conservation laws. But as regards the fundamental spatio-temporal dynamics of gravitational collapse, there was a clear and traditionally lawful way out of the apparent singularities, namely to embrace the emerging limited formalism of quantum gravity: DeWitt had plausibly derived matter bouncing back after gravitational collapse from the Wheeler-DeWitt equation \citep{dewitt_1967_quantumI}. But Wheeler now rejected this option, because his \emph{research program} told him otherwise. He held on to the tenets of his program, in particular the concept of microscopic foam-like fluctuations of topology, but conceded that these were not to be had in the framework of quantum gravity. And in thus rejecting quantum gravity, Wheeler was led to instead embrace the catastrophe of total collapse as physical and real:

\begin{quote}
Black hole squeezes out all but mass, angular momentum and charge. Collapse of universe squeezes out even these. Not 2nd quantization but higher level collapse. (NB18, p. 80, 10 November 1970)
\end{quote}

Wheeler decided that gravitational collapse was not something that needed to be fixed, but the beginning of a new era in physics:

\begin{quote}
Why stand timidly at threshold of this new domain of thought? Why not walk boldly in and exploit its richness? Is it only Columbus who shall see the gold of Mexico? (NB18, p. 124, 10 February 1971)
\end{quote}

This brings us back to the talk quoted in the introduction, which Wheeler gave in September 1972 at a symposium in Trieste celebrating Paul Dirac's 70th birthday \citep{wheeler_1973_from-relativity}. Here, total collapse was fully embraced as an unavoidable consequence of relativity: ``Today we look at the breakage that relativity has made among the laws of physics.'' (p 203). And beyond this breakage were not to be new laws, but something else:

\begin{quote}
The golden trail of science is surely not to end in nothingness. There may be no such thing as the `glittering central mechanism of the universe' to be seen behind a glass wall at the end of the trail. Not machinery but magic may be the better description of the treasure that is waiting. \citep[p. 203]{wheeler_1973_from-relativity}
\end{quote}

\begin{figure}
  \includegraphics[width=\linewidth]{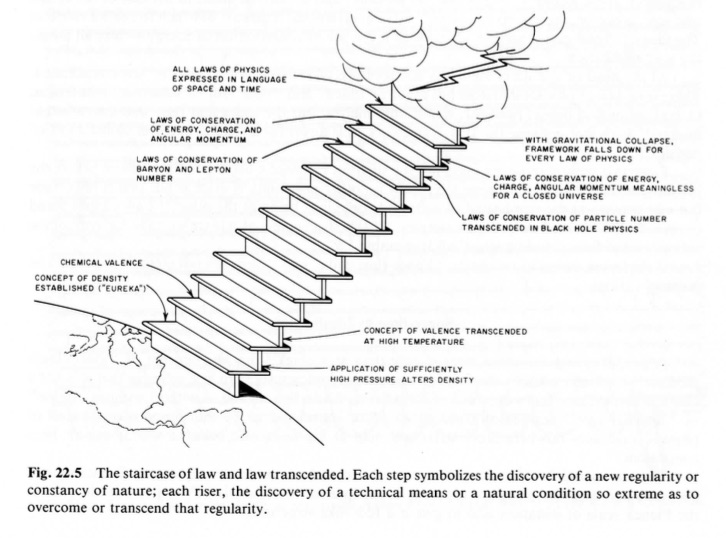}
  \caption{The Staircase of Law, and Law Transcended \citep[p. 282]{wheeler_1975_the-universe}.}
  \label{fig:staircase}
\end{figure}

We can understand Wheeler's turn away from a lawful machinery as a result of both the transcendence of the conservation laws and the fact that he was explicitly rejecting the lawful alternative of mainstream quantum gravity.\footnote{In his talk, Wheeler fielded as further reasons for lawlessness the long-recognized problematic status of energy and angular momentum in general relativity. Later he would also field the problematic status of charge in a closed universe, thus implicating all three remaining properties of a black hole in transcendence. \citep[Section 19.4]{misner_1973_gravitation}} Quantum gravity was not able to provide him with a final theory of physics, a theory that would be able not just to describe gravity up to the smallest scales, but also encompass all of particle physics. So, if a final theory was not to be had within the framework of physical law, one had to leave that framework. The last step on the ``staircase of law, and law transcended'' (Figure \ref{fig:staircase}), after ``space and time themselves are transcended as categories'' was not ultimate law, but something entirely different. But what was the relevance of the anticipated ``magic'', if it was only to reveal itself in gravitational collapse, hidden behind a horizon? To understand the full impact of the anticipated lawlessness, we need to backtrack and take a look at Wheeler's cosmology.

\section{Wheeler's Cosmology}
\label{sec:cosmo}

So far, when speaking of gravitational collapse, we have primarily considered the collapse of very massive stars. But from the very beginning, Wheeler's interest in gravitational collapse was defined also by its cosmological relevance: for gravitational collapse happens not just to stars but can, in general relativity, also happen to the dynamical universe as a whole. This process, which can be considered the inverse of the big bang is nowadays commonly referred to as the ``big crunch''.

The first question we must ask is: why did Wheeler anticipate that the universe as a whole would undergo gravitational collapse? The reason for this anticipation -- and, as we shall see, for many other elements of Wheeler's cosmology -- lies in the importance that Wheeler ascribed to Mach's principle. This principle is notoriously hard to pin down, but its core statement is rather uncontroversial: inertia is to be explained as resulting from interactions with the overall matter distribution of the universe, and not, as implied by Newtonian physics, as the effect of some absolute structures of space.

Mach’s principle\footnote{We will in the following not distinguish between Mach's original formulation (very vague) and Einstein's interpretation (still pretty vague). For a detailed analysis, cf. \citep{norton_2006_machs}.} had been an inspiration to Einstein in the formulation of general relativity and had indeed been partially integrated in general relativity: first, inertia and gravitation were unified, as both inertial motion and motion under a gravitational force were determined by the metric tensor of spacetime. Second, that metric tensor was to some extent determined by the matter distribution in the universe, making not just gravitational but also inertial motion the result of (field-mediated) interactions with the other masses in the universe.

An important question, already to Einstein, was to what extent the matter content of the universe \emph{fully determines} inertial motion, i.e., to what extent general relativity fully implements Mach's principle. If, for example, the Einstein equations are solved with asymptotically flat boundary conditions, it is really those boundary conditions that determine inertial motion, not the matter content of the universe. \citet{einstein_1917_kosmologische} thus reached the conclusion that in order to be Machian, general relativity needs to assume a spatially closed universe, in which there are no boundary conditions: if one kept on travelling through space in the same direction, one would eventually (after a \emph{very} long journey) end up where one started.

Einstein’s views on the connection between Mach's principle and general relativity as well as the cosmological conclusions he drew from it were not part of the mainstream view of general relativity. This was all the more true as Einstein's early cosmological speculations appeared to be invalidated by the discovery of the expansion of the universe. But Einstein's Machian arguments for a closed universe where really quite independent of his arguments for a static universe (to which end he had briefly introduced the cosmological constant). And Machian cosmology continued to exert a great appeal on some, in particular Wheeler; the primary reason being that Wheeler was originally envisioning a reformulation of general relativity as a theory of direct particle interactions, very much in keeping with Mach’s relationalist approach \citep{blum_2019_tokyo}. 

So strong was Wheeler's belief in Mach's principle that he included the assumption of the spatially closed universe among the axioms of general relativity in the 1958 Solvay talk, along with the statement that Einstein's second cosmological assumption of 1917 had been wrong, i.e., there was no cosmological constant, and the universe was changing in time:

\begin{quote}
In speaking about Einstein's theory [...] we mean not only the system of differential equations associated with his name, but also two further points [...]:\\
(1) The universe is closed;\\
(2) No ``cosmological'' term is to be added to the field equations.\\
It is probably best at this time to accept these points as part of what we mean by the phrase ``Einstein's version of general relativity''. \citep[p. 98]{adams_1958_some-implications}
\end{quote}

One important implication of the spatial closedness of the universe was that the universe could not go on expanding forever and would ultimately have to start contracting again.\footnote{This statement is predicated also on Wheeler's second axiom, the absence of the cosmological constant, which could, if large enough, prevent recontraction; this was of course the reason for which Einstein originally introduced it.} We can thus understand how it was Mach's principle that led to Wheeler's expectation that the universe as a whole would undergo gravitational collapse.

But while Mach's principle determined that the universe was closed in space, the question of the boundaries of the universe in time was far less obvious, and in his notebooks of the 1950s we find Wheeler mulling various options in discussion with his PhD student Charles Misner (NB3, p. 238): ``3 possibilities? Time closed [...]; time infinite; or time ends sharply``. Closed time would of course have been the most immediate complement to the spatial closure of Mach's principle, but Wheeler was doubtful of this option.\footnote{Closed time is of course a problematic notion in general; it essentially means that if you live long enough you will return to the time where you were born. But Wheeler had more specific qualms about this option, concerning the possibility of the existence of electric charges in a temporally closed universe. These qualms are not sufficiently worked out in the notebooks to be fully reconstructed, but they appear to have carried quite some weight for Wheeler.} And a  universe with sharp temporal boundaries would have brought the problem of boundary conditions back on the table. A temporally infinite universe thus seemed the most attractive possibility. But there is a difficulty here: if the universe is closed in space, it will at some point undergo gravitational collapse.\footnote{It should be noted that this conclusion is nowadays considered not as inevitable as it would have seemed to Wheeler \citep{barrow_1986_the-closed}.} Wheeler thus had to assume that this collapse was not the end of the universe, but that the collapsing universe would rebound and start expanding again; an infinite sequence of expansion and collapse; a cyclic cosmology. This was the view of the universe that Wheeler would adopt in the late 1950s and pursue all through the 1960s.

Wheeler was not the first to consider a cyclic cosmology. The general idea of course goes way back. And also in the context of general relativity, other physicists had anticipated that a possible gravitational collapse of the universe as a whole would not be the end of all things, but that the universe would somehow rebound and start expanding again \citep{kragh_2018_cyclic}. But as with the baryon non-conservation discussed in Section \ref{sec:baryon} it was one thing to state that a ``big bounce'' happened, and quite another to find a mechanism for it actually happening. Wheeler initially turned to an analogy with hydrodynamics, hints at which he discovered in a book on ``Underwater Explosions''.\footnote{It is not entirely clear what prompted Wheeler to read this book \citep{cole_1948_underwater} in the first place. The book is a scientific one, but is primarily the outcome of (and relevant for) military research on submarine warfare. Misner recalls  that Wheeler had learned about underwater explosions ``during his wartime service or in connection with it'' (Oral History interview conducted by Chris Smeenk, \url{https://www.aip.org/history-programs/niels-bohr-library/oral-histories/33697}). But Wheeler never really worked on submarine warfare; the closest thing was counseling work he did for the aircraft company Convair around 1955, which resulted in the conclusion that nuclear-powered airplanes were not feasible, as opposed to nuclear-powered submarines \citep[p. 272]{wheeler_1998_geons}. But given that the first nuclear-powered submarines had already been commissioned at the time, this work would hardly have needed original research on submarines. It is thus conceivable that Wheeler read the book either for inspiration on the hydrodynamics-GR analogy (which he explored in great detail in \citep{power_1957_thermal}) or for pleasure: Wheeler was, after all, a big fan of explosions in general, as recalled by Tullio \citet[p. 37]{regge_2012_l-infinito}: ``[Wheeler] always used to go around equipped with firecrackers and he had a true, sincere and deep passion for explosions in general. That probably explains his enthusiastic participation both in the Manhattan project and in the later project for the development of the H bomb. Just to give you an idea, from time to time he would explode devices under the university toilets. I will never forget when, one Sunday, we managed to make a lot of stuff explode in the garden of the institute... You would have to be there! To be honest, we were not expecting to create such a mess: the police quickly arrived and we had to convince the officers that we were not terrorists, as they seemed inclined to believe, but mere theoretical physicists who were studying black stars.''} While this book is primarily about the shock waves produced by such an explosion, it also contains a substantial section on the motion of the gas sphere left over by the explosion (the ``bubble''). This bubble starts contracting under the pressure of the surrounding water, but then, after being sufficiently compressed, reverses its motion again and then continues to oscillate, emitting secondary pressure waves (``bubble pulses''). The book also presented some very suggestive photographs of the bubble, showing how small departures from sphericity were amplified into horns and spikes at the point of maximal contraction. Wheeler considered this a promising ``mechanical analog for the universe'':

\begin{quote}
A look at the bubble at this time [of maximal contraction] suggests, not an ideal mathematical object, but a glove which is turning itself inside out one finger at a time. Does the universe likewise undergo repeated oscillations? show instabilities? have short-lived localized regions of high curvature and high concentrations of matter energy? have no single well-defined time of maximum contraction? \citep[p. 115f]{adams_1958_some-implications}
\end{quote}

But these hydrodynamical analogies were short-lived, as Wheeler soon started emphasizing disanalogies, most importantly that the ``universe has no internal pressure to reverse its oscillations'' (NB5, p. 252, 16 August 1958). More importantly, results started to pile up that no non-singular bounce would be possible in general relativity, culminating in Hawking's cosmological version of the singularity theorem, the analog of Penrose's result for stellar collapse \citep{hawking_1966_the-occurrence}. As Wheeler came to accept the possibility that collapse beyond the Schwarzschild horizon was possible in the early 1960s, he also accepted that quantum theory would be needed to provide a picture of how an apparent collapse to a singularity was avoided and, in the cosmological case, was replaced by a rebound \citep[p. 70]{wheeler_1962_the-universe}.

We thus see that Wheeler's assessment of cosmological gravitational collapse developed quite in parallel to his views on stellar collapse: he honed in on the problem early on, began looking for mechanisms to deal with critical aspects (baryon number non-conservation, need for a rebound), and by the late 1960s accepted that a quantum theory of gravity was necessary for a fuller description, while holding on to his basic intuitions. We will turn soon to the question of how his cosmological considerations changed when he began to turn away from his program of  geometrodynamics around 1970. But first, we need to understand how cosmological gravitational collapse was connected with the question of physical law, above and beyond the issues arising in stellar collapse that we discussed in the preceding sections.  

The central issue here was that the destruction of matter through the non-conservation of baryon number had, for Wheeler, farther-reaching implications when it occurred on a cosmological scale. The oscillations performed by the universe as a whole implied that the universe would periodically enter periods of high density, akin to those obtained in stellar collapse. In those periods, baryon number would no longer be conserved. Consequently, the universe would enter each cycle with a (possibly vastly) different matter content \citep[p. 140]{adams_1958_some-implications}. This was of course an interesting thought in and of itself.\footnote{And one that would ultimately come to be seen as making the entire setup untenable, cf. \citep[p. 257]{halpern_1995_the-cyclical}.} But the total matter content of the universe was also tied up with the laws of physics in a way that was again defined by Wheeler's reading of Mach's principle.

As we have already mentioned, Mach's principle postulated that the inertial forces were the result of interactions with distant masses. This very vague idea found some concretization in general relativity, where these interactions were to be related to gravitation; but also here, the way in which \emph{inertia} was to be understood in terms of interaction was not spelled out until 1953 when Wheeler and -- in parallel or perhaps with priority -- Dennis Sciama \citep{blum_2019_tokyo} used a schematic version of general relativity to derive Newton's law of inertia from a test body's gravitational interaction with the rest of the universe under the condition that

\begin{equation}
\label{eq:mach}
\frac{G}{c^2} \frac{M}{R} \approx 1
\end{equation}

where $G$ is Newton's gravitational constant, $c$ is the speed of light, $M$ is the total mass of the universe and $R$ its radius.\footnote{The equation in this exact form does not show up in Wheeler or Sciama, but first appears in Dicke's work \citep[p. 36]{dicke_1958_gravitation}, which we shall discuss in a moment. It can be obtained from equations in Wheeler and Sciama under some simplifying assumptions, such as uniform expansion of the universe and a homogeneous distribution of the distant matter on a spherical shell of radius $R$. The same equation had already been obtained by \citet{einstein_1917_kosmologische} -- with the numerical ratio specified to be $\pi/2$ -- but with a totally different derivation involving the cosmological constant and the assumption of a static universe.} Wheeler and Sciama initially just took this as a condition that was needed to fulfil Mach's principle, but it already clearly hinted at a close relation between the total mass of the universe and the laws obtaining within it.

Another connection between the total mass of the universe and its laws had been proposed by Dirac in 1937.\footnote{\citep{dirac_1937_the-cosmological}. See also \citep{kragh_2016_varying}.} Dirac had pointed at the connection between some very large dimensionless numbers. These large numbers were (i) the age of the universe given in atomic units\footnote{The unit of atomic time that Dirac used was the time it takes light to traverse the classical radius of the electron.}, (ii) the total mass of the universe given in proton masses, and (iii) the ratio between the electric and the gravitational forces in a hydrogen atom. The first and the third were given by Dirac to be on the order of $10^{39}$, while the second was on the order of $10^{78}$, i.e., $(10^{39})^2$. Dirac provided a tentative explanation for these correlations: both the total mass of the universe and the gravitational constant were functions of cosmological time $t$, the mass increasing as $t^2$ (presumably through some process of spontaneous creation of matter), while the gravitational constant was decreasing as $t^{-1}$, so that the gravitational force grew progressively weaker as compared to the electromagnetic one. For Dirac the temporal evolution of the universe was thus driving a change both in its total mass and in its laws, thereby leading to a correlation between the two.

These ideas were taken up in the mid-1950s by Wheeler's Princeton colleague Robert Dicke, who also brought them to Wheeler's attention. Wheeler's notebooks contain notes taken at a talk given by Dicke at Princeton on 8 November 1956 (NB5, p. 42). They clearly show Wheeler's interest, but also his concern that a varying gravitational constant would be in conflict with the geological history of earth. Wheeler remained skeptical when Dirac's hypothesis was again discussed by others at the already mentioned Solvay conference in June 1958, remarking in his conference notes: ``Conclude phys[ical] const[ant]s aren't changing.'' (NB5, p. 228). 

At around the same time, at a talk given to the Philosophical Society of Washington in April 1958, Dicke proposed an inversion of the causal relations underlying the correlations between the large numbers -- an inversion based on Mach's principle: Dicke proposed that the mass of the universe should be taken as the primary large number. It determined the smallness of the gravitational constant through something like the Mach-inspired equation \ref{eq:mach}. The problem was then of course to explain how the age of the universe could be determined by the other large numbers; after all, the age of the universe seemed to be a brute fact if there ever was one. Here Dicke introduced what would later be called ``anthropic'' arguments:

\begin{quote}
The present epoch is conditioned by the fact that the biological conditions for the existence of man must be satisfied. This requires the existence of a planetary system and a hot star. \citep[p. 33]{dicke_1958_gravitation}
\end{quote}

Somehow, Dicke argued, the time needed to achieve the right ``biological conditions'' would depend on the size of the gravitational constant, thereby establishing the second causal link needed to explain the large number correlations. But it was not these anthropic considerations that initially appealed to Wheeler, although they would become, as we shall see, important later on. Rather, it was the idea of using Mach's principle not just to explain inertia and to deduce the closedness of the universe, but also to interpret natural constants as functions of the mass of the universe. In discussions with Dicke (NB5, p. 284, 24 September 1958), Wheeler remarked that Mach's principle might ``change masses of elem[entary] particles or fine structure constant.''

This idea came to new life after Wheeler in the early 1960s embraced the conclusion that the big bounce would be a genuinely quantum-mechanical process. Combined with the fact that baryon number (and thus the total mass of the universe) was not conserved in gravitational collapse, this implied that the initial matter content of a given cycle of the universe was merely determined probabilistically, as were then the natural constants determined through Mach's Principle. Wheeler began exploring this idea, evoking geological metaphors that recalled the original difficulties involved in having constants varying with time:

\begin{quote}
We are accustomed to the idea of fossil molecules and fossil atomic nuclei. Should we also accustom ourselves to the idea of fossil elementary particles? The piece of wood on which I lay my hand was made in a tree a few years ago from carbon dioxide and water. Its molecules are ``chemical fossils'' of recent origin. The atomic nuclei of these carbon and oxygen atoms date back to an older epoch some billions of years ago the time of thermonuclear combustion in the stars. They are ``nuclear fossils''. In another chemical environment the carbon dioxide and the water would have joined into a molecule quite different from cellulose. Nuclear matter cooked in a different star for a different length of time would have left us more iron and less oxygen as its fossil testament. Are we likewise to expect that the properties of the elementary particles and even their numbers differ between one cycle of the universe and another? \citep[p. 268-269]{wheeler_1967_our-universe}
\end{quote}

But he was still looking for a mechanism to explain how exactly the essentially random initial conditions determined the observed properties of the elementary particles, a question not answered by the invocation of Mach's principle. He recalled Laplace's Demon, an idea he still essentially considered adequate, remarking only that ``[t]he reader of 1814 [...] was wrong if he thought one would uncover the rationale of the initial conditions'' (p. 271). The subsequent notebooks thus contain numerous attempts at trying to determine ``how can size of universe affect properties of elementary particles'' (NB 15, p. 87, 8 July 1968). He considered an analogy with the geodesic dome of Buckminster Fuller, where the size and shape of the dome at large determines the number of pentagons it was composed of, ``a line of comparison that might almost be called Keplerian!'' He explored the possibility that the identical particles observed today would only gradually emerge in the evolution of the universe:

\begin{quote}
I have argued that [particle] properties are not superuniversal but differ from one history of the universe to another. Therefore a particle building phase must occur. Up to now I have always assumed that this phase is to be identified with the moment when the universe has the dimensions $L^{\ast} = (\hbar G/c^3)^{1/2}$. However, I see [...] that there is another possibility: that \emph{the particles are not formed until sometime after the universe has commenced its expansion. Time} is needed to ensure that the various ``bubbles in the dough'' all have the same size. (NB 15, p. 174, 10 November 1968, emphasis in the original)
\end{quote}

But everything changed with the fall of geometrodynamics discussed in the previous section. We have now finally reached the point where we can analyze the impact of Wheeler's embrace of lawlessness on his cosmology and then in turn analyze the implications of cosmology for lawlessness. We have identified and traced the origins of several central elements in Wheeler's cosmological thought in the late 1960s, just before the fall of geometrodynamics: the idea of a cyclic cosmology; the impact of cosmology, and in particular of the initial conditions of a given cycle, on elements of physical theory that would normally be considered immutable, in particular the mass spectrum of the elementary particles; and the search for a mechanism that would ensure this connection between the cosmological and the microphysical. With the abandonment of geometrodynamics it was now no longer just the matter content of the universe (and the natural constants determined by it) that was wiped out in the gravitational collapse of the universe; it was joined by ``everything that one has ever called a law of physics.'' In Wheeler's cosmological thinking, this of course implied that all laws of physics were now to be considered cosmologically determined. But it also meant that there was no point in searching for a mechanism for how this determination occurred, because there were no more ``superuniversal'' laws that such a mechanism could obey.

Of course, the ``golden trail of science'' did not end there. In the absence of laws to argue from, Wheeler instead began to invoke ``the quantum principle'', a term he would continue emphasizing (sometimes simply as ``the quantum'') until the end of his days, in the new millennium \citep{tegmark_2001_100}. What did it stand for? One might at first be tempted to believe that Wheeler was not actually embracing lawlessness, that he had merely rejected relativity or quantum gravity, and was now instead pursuing a new daring conservatism based on the established framework of quantum mechanics or quantum field theory. This is indeed how he was interpreted, e.g., by Steven Weinberg:

\begin{quote}
My friend and teacher John Wheeler has occasionally suggested that there is no fundamental law and that all the laws we study today are imposed on nature by the way that we make observations. (…) Wheeler (…) simply seem(s) to me to be merely pushing back the problem of the final laws. Wheeler’s world without law still needs metalaws to tell us how our observations impose regularities on nature, among which metalaws is quantum mechanics itself. [...]
I expect that all attempts to do without fundamental laws of nature, if successful at all, simply result in the introduction of metalaws that describe how what we now call laws came about. \citep[p. 186]{weinberg_1993_dreams}
\end{quote}

This does not do justice to Wheeler's invocation of the ``quantum principle''.\footnote{That is not to say that Wheeler was not impressed by the contemporary developments in quantum field theory. The early 1970s was also the time of the establishment of the Standard Model of Particle Physics \citep{hoddeson_1997_the-rise}, and this development was clearly a further influence in Wheeler's realization that geometrodynamics was an insufficient foundation for all physical theory: ``general relativity [...] created a new standard for the surprise of a prediction and for the scope of an explanation. The standard has meantime risen, not least because of the beautiful regularities uncovered in particle physics. General relativity has not kept up with the rise.'' \citep[p. 236]{wheeler_1973_from-relativity}} The expression was Wheeler's and it did not refer to any sort of metalaw, like the least action principle, say, from which any part of the formal apparatus of quantum mechanics could be derived directly. Wheeler was referring to a far more abstract, philosophical core idea of quantum physics: the relation between observer and observed system, which Wheeler had long viewed as one of the central lessons of his mentor, Niels Bohr \citep[p. 30]{wheeler_1963_no-fugitive}. While the quantum principle was thus grounded in history, it was also transformed by Wheeler:\footnote{As is typical of Wheeler's use of history, see \citep{furlan_2020_einsteins}.} unlike Bohr's observer, Wheeler's observer was to be of cosmological and cosmogonic significance. Wheeler explored this significance in a series of talks he gave in the mid-1970s, tellingly entitled ``The universe as home for man'' \citep{wheeler_1975_the-universe};\footnote{Wheeler presented this talk at a symposium celebrating the 500th birthday of Copernicus. The irony of presenting an anthropocentric cosmology on this occasion was not lost on the participants, with the bacterial physiologist Bernard Davis remarking that it was ``a little bit contrary to the Copernican tradition''. \citep[p. 582]{gingerich_1975_the-nature}} ``Is physics legislated by cosmogony?'' \citep{patton_1975_is-physics};  and ``Genesis and observership'' \citep{wheeler_1977_genesis}.

The question of the observer had played a role in Wheeler's cosmological thought early on, as it raised the question of how to interpret the wave function of the universe \citep{blum_2022_the-interpretation}. An extrapolation of the Copenhagen interpretation, with its dichotomy between the classical observer and the observed quantum system, appeared to require an observer with a god-like point of view, looking at spacetime from the outside, the system under observation being the universe itself. This was a problematic conclusion, to say the least. The view proposed by Wheeler's 1957 PhD student Hugh Everett III offered an attractive alternative: observers are just subsystems \emph{within} the world. Everett described these subsystems in his thesis draft as ``automatically functioning machines, possessing sensory apparatus and coupled to recording devices capable of registering past sensory data'' (reprinted in \citep[p. 64]{everett_1973_the-theory}). Wheeler preferred to think of them as ``sufficiently complex to simulate all that human beings do, including acting, observing, and recording'' and as having ``arisen within the overall system by organic evolution'' (Wheeler to Alexander Stern, 25 May 1956, reprinted in \citep[p. 221]{barrett_2012_the-everett}).

After the rejection of Everett's ideas by Bohr and other Copenhagen physicists \citep{osnaghi_2009_the-origin}, Wheeler left the problem of the embedded observer more or less alone for a decade,\footnote{Though it did not disappear entirely from his thinking, cf. \citep[p. 187-188]{wright_2014_more-than-nothing}.}  before returning to it in the early 1970s in the wake of the fall of geometrodynamics. His take was rather unorthodox, but this should not be taken to imply that he was positioning himself against the Copenhagen interpretation, or in any way positioning himself within the emerging ``interpretation debate''.\footnote{On which, see \citep{freire_2015_the-quantum}.} ``Interpretations'' such as Everett's were just one of many ways of approaching the quantum principle (all potentially fruitful), on par with Feynman's path integral or the quantum logic of Birkhoff and von Neumann \citep[p. 288]{wheeler_1975_the-universe},\footnote{In view of the spectrum of alternatives that has crystallized into today’s interpretation debate, there is a quite striking absence in the many facets of the quantum principle invoked by Wheeler: Bohmian mechanics. While a comprehensive story of the relationship between Wheeler and David Bohm, also on a personal level, has still to be written, it seems safe to state here that such approaches were excluded by Wheeler quite simply because they pointed in a direction exactly opposite to his own, by trying to restore a classical level beneath the quantum.} a principle so many-faceted it was admittedly hard to grasp:

\begin{quote}
Quantum principle? What an inadequate name for an overarching feature, or \emph{the} overarching feature, of nature. We understand any other principle of physics in enough completeness to summarize it, beginning with a good name, in a dozen words. But not this. It continually unfolds with fresh meaning. It might almost better be called the ``Merlin principle''. Merlin the magician, pursued, changed first to a fox, then to a rabbit, and finally to a bird fluttering on one's shoulder.\citep[p. 287]{wheeler_1975_the-universe}
\end{quote}

Indeed, Wheeler's take on the quantum principle is best understood as a combination of several elements: Bohr's complementarity, Everett's ideas on the embedding of the observer, and also the views of his Princeton colleague Eugene Wigner on the role of human consciousness in measurement (on which more in the next section). The resulting statement of the quantum principle was ultimately very simple: 

\begin{quote}
[B]eyond the rules of quantum mechanics for calculating answers from a Hamiltonian stands the quantum principle. [...]  It promotes observer to participator. It joins participator with system in a ``wholeness'' (Niels Bohr) [...] quite foreign to classical physics. It demolishes the view we once had that the universe sits safely ``out there'', that we can observe what goes on in it from behind a foot-thick slab of plate glass without ourselves being involved in what goes on. We have learned that to observe even so miniscule an object as an electron we have to shatter that slab of glass. We have to reach out and insert a measuring device. We can put in a device to measure position or we can insert a device to measure momentum. But the installation of one prevents the insertion of the other. We ourselves have to decide which it is that we will do. Whichever it is, it has an unpredictable effect on the future of that electron. To that degree the future of the universe is changed. We changed it. We have to cross out that old word ``observer'' and replace it by the new word ``participator''.\footnote{\citep[pp. 560-562]{patton_1975_is-physics}} 
\end{quote}

And the influence of this ``participator'' could also stretch back in time, e.g., when observing ``radiative evidence of what went on in the first few seconds after the big bang'' \citep[p. 25]{wheeler_1977_genesis}.\footnote{This observation can be viewed as an anticipation of Wheeler's later work on cosmological delayed-choice experiments \citep[p. 190]{wheeler_1983_law}.} Without the observer, possibilities do not become actual existences. It was this ability of observers to actualize possibilities\footnote{An element of quantum theory for which Wheeler liked to invoke a quote by William James, ``Actualities seem to float in a wider sea of possibilities from out of which they are chosen'', e.g., in \citep[p. 288]{wheeler_1969_from-mendeleevs}.} on cosmic scales that allowed them to replace the mechanism for determining the laws of the universe that Wheeler had been looking for. Going even further, it allowed the observer to be responsible for the genesis of the universe itself, to be a ``participator in genesis'' \citep[p. 25]{wheeler_1977_genesis}. No endless cycles of the universe with probabilistic initial conditions were needed anymore; the preference for cyclic cosmology had anyway, as we have seen, been rooted in a very lawful reading of general relativity and Mach's principle.\footnote{It should be noted, however, that Wheeler did not abandon his Mach-inspired views on the \emph{spatial} boundary conditions of the universe and continued to assume that the universe was closed. He thus did not question the possibility that the universe would end in a big crunch, and his cosmology was thus still heavily shaped by Mach's Principle.} In Wheeler's new ``self-reference cosmology'' observership was sufficient for explaining cosmogony:

\begin{quote}
Self-reference cosmology has these features: (1) one cycle only; (2) the laws and constants and initial conditions of physics frozen in at the big bang that brings the cycle and dissolved away in the final extremity of collapse; and (3) a guiding principle of ``wiring together'' past, present and future that does not even let the universe come into being unless and until the blind accidents of evolution are guaranteed to produce, for some non-zero stretch of time in its history-to-be, the consciousness, and communicating community, that will give \emph{meaning}\footnote{Here Wheeler included a footnote, quoting the Norwegian philosopher Dagfin F\o llesdal: ``Meaning is the joint product of all the evidence that is available to those who communicate.''} to the universe from start to finish \citep[p. 566-567]{patton_1975_is-physics}
\end{quote}

This then was Wheeler's re-imagination of Dicke's anthropic arguments. We might end out story of how John Wheeler lost his faith in the law here. We have seen how the transcendence of law in gravitational collapse and the apparent inability of physics to provide a satisfactory final theory led Wheeler to abandon a traditional view of reductionist physical law and instead embrace the view that on a cosmological level ``the observer himself closes up full circle the links of interdependence between the successive levels of structure'' \citep[p. 3]{wheeler_1977_genesis}. We have also seen that this invocation of the observer can hardly be considered, as Weinberg did, as falling back on (the metalaws) of quantum mechanics.

 That is not to say that there is no element of ``pushing back of the problem'' in Wheeler's post-geometrodynamical thinking. But it was pushed back out of physics and into different fields of thought. When exploring the structure of what he referred to as ``pregeometry'', Wheeler started drawing on mathematics and propositional logic, culminating in his slogan ``it from bit'' \citep{wheeler_1990_information}.\footnote{See \citep[Section 4.5]{wright_2014_more-than-nothing}. It should also be note that Wheeler's turn to propositional logic was also related to the Merlin principle, another facet of which was the quantum logic of Birkhoff and von Neumann.} This would bring us to the relation between laws of nature and the laws of mathematics and logic, which is beyond the scope of this paper. But there is another aspect of Wheeler's pushing back that relates more directly to laws of nature, in particular to laws from other disciplines of the natural sciences. For while Wheeler's notion that the observer is responsible, as it were, for the creation of universe might at first glance sound like naive idealism, he was not claiming that the universe is a projection or creation of the mind. Mind could not be primary, because it, in turn, had to be brought into being by the universe. The question of how mind (and with it life) were constituted and how they came into being was thus an important element of Wheeler's self-reference cosmology. In the last section we will thus offer a brief gloss of Wheeler's engagement with the laws of (evolutionary) biology.

\section{Creating the Observer}
\label{sec:bio}

Wheeler was not the only one in the early 1970s thinking about what constituted the observer. In NB18, p. 219, 10 June 1971, in a small section quite significantly entitled ``Program'', we find a reference to a recent work by \citet{wigner_1970_physics}, ``Physics and the explanation of life''. In the 1960s, Wigner had gotten interested in foundational questions of quantum mechanics and had put forth his ideas that human consciousness might be responsible for the collapse of the wave function \citep[Chapter 4]{freire_2015_the-quantum}. In the 1970 paper, Wigner observed

\begin{quote}
[T]here is a tendency in both physics, which we consider as the most basic science dealing with inanimate objects, and in the life sciences to expand toward each other. Furthermore, the tendency is strongest in the modern parts of the two disciplines: in quantum mechanics on the one hand, in microbiology on the other. Each feels it cannot get along by relying solely on its own concepts. (p. 39)
\end{quote}

Wigner did not venture to propose some specific idea (or mechanism) to bridge the gap. But he did present the two main alternatives: either the ``laws of nature, for the formulation of which observations on inanimate matter suffice, are valid also for living beings'' (p. 39)\footnote{Though it is important to specify here that “this assumption need not imply, as is often postulated, that the mind and the consciousness are only unimportant derived concepts which need not enter the theory at all” (p. 41).} or ``the laws of physics will have to be modified drastically if they are to account for the phenomena of life'' (p. 41). In the paper, Wigner clearly expressed his preference for the second option. 

Which brings us back to Wheeler, whose program was rather ambiguous regarding this point. On the one hand, the physical laws, once constituted in the big bang, were supposed to give rise by themselves to the conscious observer. This was in keeping with his earlier daring-conservative methodology, which might be called reductionist, in the sense of starting from fundamental microscopic laws, while also being exploratory, an open-ended pursuit of the manifold possibilities those fundamental laws provided. Viewed from this angle, invoking a modification of these fundamental laws for the purpose of obtaining consciousness would appear as a ``free play with ideas'' and thus as the opposite of daring conservatism.

On the other hand, however, in the cosmological reflections Wheeler was developing, physical law and the existence of the universe were based on the existence of sentient observers, a view that seems far closer to the second option. Wheeler thus concurred with Wigner that ``[i]f the concept of observation is to be further analyzed, it cannot play the primitive role it now plays in the theory'' (p. 43). In the early 1970s, consequently, Wheeler started reaching out to scientists from beyond physics, who might be able to proved insights for fleshing out the concept of observer in his new anthropic cosmology. For instance, he sent ``The Universe as Home for Man'' to the neurophysiologist John Eccles. Eccles seemed pleased with the paper (Letter to Wheeler of 30 December 1975, Wheeler Papers, Box 8), which appeared to resonate with his own, more philosophical, work in which he argued against the ``materialist monists'' \citep{eccles_1977_the-self}. Not much of a dialogue, however, seems to have been born from this brief exchange. Richer and more fruitful were the interaction that Wheeler had with two eminent scientists who, like Wheeler, had a background in the physical sciences, but had become interested in questions of life and even consciousness, Manfred Eigen and Ilya Prigogine.

Eigen’s talk at the 1972 Trieste Symposium (where Wheeler had first enunciated the end of all physical law) made a great impression on Wheeler. A couple of years later Wheeler wrote to Eigen that ``[i]n the Trieste seminar of two years ago [...] no talk more captured my imagination, no talk seemed more to break new ground, and none opened wider vistas than your own'' (Wheeler Papers, Box 8). Eigen was then invited to Princeton -- to give a colloquium on, Wheeler suggested, ``Physical and Chemical Background of the Origin of Life'' and for ``a Trieste reunion'' with ``[a]nother one of our Trieste symposium colleagues, Eugene Wigner''. Let us try, then, to get an idea of what Wheeler appreciated in Eigen’s talk and why he was so keen on exchanging ideas with someone who “belong[ed] just as much to the world of physics as to the world of biology”. 

Eigen had begun his Trieste talk by stating: “In our days the physicist's conception of nature certainly has to include the phenomenon of life”. Eigen went on to illustrate his hypercycle model for the origin of life,\footnote{By using systems of non-linear differential equations, Eigen considered a class of self-replicating macromolecules (possibly nucleic acids — the topic of the speech, after all, concerned life and its origins), linked in such a way that each one of them acts as a catalyst for the formation of the next one, with the last one catalyzing the first. "Such a cyclic hierarchy of catalytic processes may be called a `hypercycle''' \citep[p. 623]{eigen_1973_the-origin}.} not as an ultimate answer, but as a possible physical mechanism worthy of exploration:

\begin{quote}
Do the known laws of physics provide a complete basis for an understanding of the phenomenon life? All we can do at the moment is, to disprove the negation of such a claim by quoting counter-examples. In other words we can use physical models to disprove any claim that the known laws of physics are not sufficient to describe the phenomena which specify a living object. The hypercycle is one example. \citep[p. 631]{eigen_1973_the-origin}
\end{quote}

In tentatively exploring what we have referred to as Wigner’s first option, Eigen placed an emphasis on known laws -- a heuristic strategy that is clearly reminiscent of daring conservatism. Before making definitive proclamations or invoking the need for radical changes, Eigen adopted and insisted on a constructive attitude that tried to squeeze out the potential inherent in established laws. Eigen was offering a picture where some underlying randomness became encapsulated in more and more complex levels of organization, through some sort of selection process. Even if the actual dialogue between Wheeler and Eigen appears, at least in the light of currently available archival material, not to have been that extended, Wheeler would go on to prominently cite Eigen's work at the beginning of his 1977 Varenna lectures, which represented the first sum of Wheeler's post-geometrodynamical thinking and introduced the phrase ``law without law'' \citep{wheeler_1979_frontiers}. Here Eigen was (the only one!) credited with having shown that ``[t]he hierarchical speciation of plant and animal life [...] arises out of the blind accidents of genetic mutation and natural selection'', i.e., that evolution proceeding in accord with known physical laws could indeed bring forth an observer.\footnote{Eigen, in turn, would quote the grand finale of Wheeler's Trieste talk, ``We can believe that we will first understand how simple the universe is when we recognize how strange it is'' in his book ``Das Spiel. Naturgesetze steuern den Zufall'' \citep[p. 244]{eigen_1978_das-spiel}.}

Prigogine's ideas were more radical, advocating the need for a fresh conceptual reformation at the basis of physics, closer to Wigner's second option. These ideas resonated with Wheeler's post-geometrodynamical thinking. Like Eigen, Prigogine had attended the Trieste symposium, where he had given two talks, the main one being on ``Time, Irreversibility and Structure'', where he, too, spoke on how to get structure from randomness, in this case (quantum)\footnote{Prigogine had worked \citep{george_1972_the-macroscopic} on the foundations of quantum mechanics, co-authoring a paper with L\'eon Rosenfeld, in which they aimed to refine Bohr’s ideas on measurement  -- since, as Prigogine later stated, ``Bohr’s ideas were largely intuitive. Bohr has said deep things, but in a rather obscure fashion'' \citep[p. 75]{buckley_1979_a-question}. Indeed, his second, shorter talk at Trieste was on the subject of ``Measurement Process and the Macroscopic Level of Quantum Mechanics'' \citep{prigogine_1973_measurement}.} fluctuations:

 \begin{quote}
Now fluctuations - the spontaneous deviations from some average regime - are a universal phenomenon of molecular origin and are always present in a system with many degrees of freedom. [...]\\
a new order principle appears that corresponds essentially to an amplification of fluctuations and to their ultimate stabilization by the flow of matter and energy from the surroundings. We may call this principle 'order through fluctuations'. \citep[p. 587]{prigogine_1973_time}
\end{quote}

But their main interactions occurred after Wheeler retired from Princeton and moved to the University of Texas in 1976, where Prigogine was a professor.\footnote{Prigogine highlighted his interactions with Wheeler several years later in \citep{geheniau_1988_the-birth}.} Prigogine's work on non-equilibrium thermodynamics had offered him the hints for his new vision: complex ``dissipative'' structures, self-organized in (local) defiance of the second law of thermodynamics, had a characteristic lifetime within which their complexity was built and ultimately dissolved; no view of time as a fourth dimension or as a mere parameter that "flows" could do justice to this \emph{creative} role of time, an explicitly Bergsonian notion. Mankind's creativity is thus not an accident that developed meaninglessly in the midst of barren immensities. This was a fundamental insight underlying Prigogine's proclamation of a ``Nouvelle Alliance'' \citep{prigogine_1979_la-nouvelle} between scientific and humanistic knowledge, working together to provide a picture of the universe where we have a place and can find a meaning.
 
Even from this cursory outlook, the affinities with the vision Wheeler began to develop in the early 1970s on are evident, as are the common topics. One commonality lies in their view of statistical quantum fluctuations not as undesirable deviations from a given order, but as the basis out of which order and complexity can emerge (quantum foam,  order through fluctuations).\footnote{Wheeler would use similar wording when speaking about Darwin: ``It's inspiring to read the life of Charles Darwin and think how the division of plant and animal kingdoms, all this myriad of order, came about through the miracles of evolution, natural selection and chance mutation. To me this is a marvelous indication that you can get order by starting with disorder.'' \citep{wheeler_1979_from-the-big}}  Wheeler  himself would highlight the analogies between Prigogine's dissipative structures and his universe, which was similarly self-organized (lawless) and built and ultimately dissolved complexity (the observer) within its characteristic lifetime from big bang to big crunch \citep[p. 154]{wheeler_1979_frontiers}.\footnote{Prigogine, in turn, would approvingly refer to Wheeler's observer-participator, highlighting it, e.g., in later and updated translations of \emph{La Nouvelle Alliance} \citep[p. 267]{prigogine_1981_dialog}. He would also frequently refer to Wheelerian themes, including a chapter on irreversibility and spacetime structure in \citep{prigogine_1985_vom-sein} or an appendix on black holes in \citep{nicolis_1989_exploring}.} But this juxtaposition also revealed a fundamental disanalogy, which Wheeler also highlighted: for the universe the end in a big crunch is pre-ordained, and we also have \emph{final} boundary conditions. For Prigogine, and even more strongly for Eigen, the resolution of the dichotomy between physical matter on the one hand and life (and, for Prigogine, even meaning) on the other is achieved (or achievable) on a local level, where the laws of physics could be complemented with novel mechanisms (hypercycles, dissipative structures) and (in Prigogine's case) new conceptualizations of time to produce life and all that comes with it. Wheeler's vision, instead, was global and made reference to the whole universe. It was in the creation and in the collapse of the universe as a whole that the foundation in physical law vanished, a foundation that remained central to Eigen's and Prigogine's thinking, serving them (as it had Wheeler before the fall of geometrodynamics) as a resource for mechanism. Wheeler, on the other hand, had to do without law and thus had to rely on a form of teleology.

This is precisely where Max Delbrück put his finger. Delbrück, who had turned from physics to biology many years earlier, influenced by Bohr’s speculations on the connection between life and quantum mechanics,\footnote{See, e.g., \citep{delbruck_1949_a-physicist}.} had received from Wheeler a copy of ``Is Physics Legislated by Cosmogony?''. In his reply thanking Wheeler, Delbrück raised some objections to the way evolution was being invoked:

\begin{quote}
What bothers me with respect to your universe which self-excites into the development (evolution) of mind is this: it ignores some facts we know about the evolution of mind, viz., that the brain (and mind) evolved in \emph{adaptation} to very specific requirements for getting along and permitting its possessor to survive, in the water, on the ground, in the trees, in the cave. [...] It did \underline{\emph{not}} evolve in adaptation to a requirement to invent relativity, Qu[antum] M[echanics], cosmogony, etc.\footnote{Letter from Delbr\"{u}ck to Wheeler, 14 April 1975; Max Delbr\"{u}ck Papers, California Institute of Technology Archives and Special Collections, Box 24, Folder 10.}
\end{quote}

In other terms, the cosmogonic role of the observer introduces a teleology which is in contrast with the observer as product of Darwinian evolution, with all its accidents and contingencies; yet both aspects were supposed to be present in Wheeler's cosmology. Wheeler's answer is not extant, but an answer of sorts, albeit a paradoxical one, can be found in the very paper he had sent to Delbr\"{u}ck. Wheeler was clearly genuine in wanting to accept the role of the environment and of chance in the genesis and evolution of more and more complex structures. He was, as we have seen, also receptive to attempts at providing, on the basis of modern physics, an intrinsically random mechanism for Darwinian evolution to proceed on; and, perhaps, at providing a physical description of life and its origins, as suggested -- at least in terms of viability -- by Eigen and Prigogine. But on the global and cosmogonic level that Wheeler was trying to face, that alone could not be enough. That was why in \citep{patton_1975_is-physics} he had appealed to a deep ``wiring together'' of past, present and future and invoked the following colorful metaphor for combining teleology and random evolution:

\begin{quote}
Chance mutation, yes; Darwinian evolution yes; yes, the general is free to move his troops by throwing dice if he chooses but he is shot if he loses the battle. Deprived of all meaning, stripped of any possibility to exist is any would-be universe where Darwinian evolution brings forth no community of evidence-sharing participators... (p. 567)
\end{quote}

This does not exactly sound like a scientific justification; it sounds more like a half-ironical play with ideas. Wheeler was clearly struggling to bring across the radical conclusions he had drawn from the long and open-ended intellectual development we have been reconstructing in this paper.\footnote{Already in 1967 we find him lamenting in his notebooks: ``Why is everyone so silent? Why does it have to me who says this? Story \emph{crying} to be told. (NB 14, p. 140, 27 January 1967)} Nevertheless, the idea remained problematic, to use a slight euphemism, and the tension pointed out by Delbrück was there to stay: Wheeler's invocation of a “Darwinian” mechanism, regardless of its convoluted paths, \emph{had to} give rise to the observer -- which appears incompatible with a Darwinian world view. What we see emerging here is -- it should be noted -- a \emph{new} kind of global tension between biology and physics that went beyond any local reconciliation.

This issue was not resolved. Instead, Wheeler's observer would be gradually de-anthropomorphized. Originally his observers had been very human. But the specifics of this observer moved to the background. Once in Texas, Wheeler's focus shifted to the process (rather than the agent) of quantum measurement \citep{wheeler_1983_quantum}. Already in the Varenna lectures we can notice such a tendency. After remarking that the “strange necessity of the quantum as we see it everywhere in the scheme of physics comes from the requirement that—via observer-participancy—the universe should have a way to come into being”, Wheeler highlights the prerequisite for the cosmic game (a metaphor, not unlike Eigen’s, that runs through all the Varenna lectures) to start: 

\begin{quote}
I know that in that empty courtyard many a game cannot be a game until a line has been drawn—it does not matter where—to separate one side from the other. [...] But how much arbitrariness is there in this more ethereal kind of demarcation, the line between ``system'' and ``observing device''. \citep[p. 18]{wheeler_1979_frontiers}. 
\end{quote}

Instead of stressing the human features of the observer (or consciousness), Wheeler adopted a much more sober, in a sense even minimalistic, view:

\begin{quote}
The same stick, when grasped firmly and used to explore something else,  becomes an extension of the observer or—when we depersonalize—a part of the measuring equipment. As we withdraw the stick from the one role, and recast it in the other role, we transpose the line of demarcation from one end of it to the other. The distinction between the probed and the probe, so evident at this scale of the everyday, is the without-which-nothing of every elementary phenomenon, of every "closed" quantum process.\\
Do we possess today any mathematical or legalistic formula for what the line is or where it is to be drawn?\\
No.\\
Then what is important about this demarcation?\\
Existence, yes; position, no.
\citep[pp. 18-19]{wheeler_1979_frontiers}
\end{quote}

The distinction between the probed and the probe, together with an explicit rebuttal of the centrality of consciousness, reappears in paper very explicitly entitled ``Not Consciousness but the Distinction Between the Probe and the Probed as Central to the Elemental Quantum Act of Observation'' \citep{wheeler_1981_not-consciousness}. Wheeler was now exploring a different facet of the quantum principle:

\begin{quote}
What one word does most to capture the central new lesson of the quantum? `Uncertainty,' so it seemed at one time; then `indeterminism'; then `complementarity'; but Bohr's final word `phenomenon' – or, more specifically, `elementary quantum phenomenon' – comes still closer to hitting the point. [...] In today's words, no elementary quantum phenomenon is a phenomenon until it is a registered (`observed' or `indelibly recorded') phenomenon, `brought to a close' by an `irreversible act of amplification'. \citep{miller_1984_delayed-choice}\footnote{On Bohr's use of the term phenomenon, cf. \citep[fn. 91]{wheeler_1979_beyond}.}
\end{quote}

There was clearly a strong shift of emphasis, even if Wheeler never rejected his previous anthropic considerations.\footnote{He would write, e.g., the foreword to John Barrow and Frank Tipler's ``The Anthropic Cosmological Principle'' \citep{barrow_1986_the-anthropic}.} Even if humankind was no longer the protagonist of his worldview, humankind was still very much his intended audience. Not unlike Prigogine, Wheeler kept trying to evoke a worldview where humans could feel “at home in the universe” (1994) and be reassured of its harmony \citep{wheeler_1994_at-home}.\footnote{Wheeler's hopes for the effect this reassurance would have on humanity could be, to put it mildly, rather idiosyncratic. As he remarked in some of his first notes on anthropic arguments: ``What strange effect would it have on everyone's morality, our resolution in the war in Viet-Nam and everywhere else, our image of ourselves and our children, to have such views accepted.'' (NB14, p. 173f, 3 June 1967)} Indeed, despite affirming the lawlessness of the universe, the resulting picture is not at all one of chaos and meaninglessness, but that of a cosmos where a sort of pre-established harmony (a concept for which Wheeler drew on a tradition that he traced back via Einstein to Leibniz \citep{wheeler_1983_einsteins}) arises as the background for the cosmic game.

\section{Conclusions}
\label{sec:6}

We have followed John Wheeler's intellectual trajectory on the long (and sometimes winding) path from a resolute belief in the laws of physics to an embrace of fundamental lawlessness and the search for new ways of constituting the physical world using concepts from the life sciences. Let us briefly recapitulate: In the late 1960s, Wheeler realized that the status of physical laws seemed to be modified in gravitational collapse. Laws were being ``transcended'' (Section \ref{sec:baryon}). Around the same time, his hopes for constructing a theory of everything from the foundations of current physical theory (general relativity and quantum mechanics) faltered (Section \ref{sec:geometro}). These two factors led him to embrace ideas of fundamental lawlessness. Wheeler attempted to fill the resulting void through a novel combination of cosmology and consciousness, his participatory universe. This brought in questions and concepts from biology on two levels, through the constitution of the live and sentient observer and through the need of invoking a deep-time evolution that combined the evolution of the cosmos as a whole with the biological evolution of the observer. 

These ideas remained vague and problematic. Ultimately, the fundamental lawlessness that Wheeler embraced when conferring a cosmogonic (and nomogonic) role to his observer-participator (Section \ref{sec:cosmo}) provided no clear basis for further scientific investigation. His move away from an anthropic observer in the second half of the 1970s (see Section \ref{sec:bio}) also implied a mitigation of the radical lawlessness that we highlighted in this paper. In this later phase (bearing in mind the difficulties of periodizing Wheeler mentioned in the introduction), Wheeler entered for the first time the grey zone between the strict lawfulness of daring conservatism and the total lawlessness of the cosmogonic observer-participator, the grey zone where underlying basic principles are intertwined with the emergence of statistical regularities from a large number of individual events. In this sense, Wheeler over time traversed the entire spectrum of alternatives that defines the debate on laws of nature to this day. But what our historical reconstruction hopefully makes clear is that this was not merely the result of changes in philosophical attitude; it was rather Wheeler’s methodology of pushing ideas to their limits -- and the novel physical insights that it brought -- which led Wheeler, almost dialectically, from a position to another. Similarly, the radical lawlessness he explored might seem whimsical,  a ``postmodern'' play with ideas,  when looking at some of Wheeler's writings in isolation. But Wheeler's cross-disciplinary reconfiguration of ideas was not something that he imported from wider cultural contexts; it arose from a crisis within modern fundamental physics itself, where laws were transcended and appeared insufficient to ground all of physics by themselves.

 It should also be remarked that this crisis was one that did not result directly from the great conceptual upheavals of early 20th century physics; in spite of their epistemological novelty, relativity and quantum mechanics had been well integrated into and ultimately even strengthened -- also in Wheeler's thinking -- a lawful (though non-deterministic) view of the physical universe by vastly extending the scope of scientific explanation. It was rather later developments, related to the attempts at combining relativity and quantum theory, that, for Wheeler, set in motion the crisis of law.

Wheeler’s turn to radical lawlessness was not just unsustainable (at least if one wanted to continue doing science, as Wheeler evidently did), it may well also have seemed premature at the time. After all, the 1980s saw the rise of string theory, which promised to be both a quantum theory of gravity and a theory of everything, thereby fulfilling Wheeler's ambitions for geometrodynamics.\footnote{Wheeler's only assessment (that we are aware of) of string theory can be found in his autobiography: ``intriguing'' \citep[p. 320]{wheeler_1998_geons}.} At the same time, advances were made in ``quantum gravity proper'', which appeared to imply that such a theory might have more to say about the structure of spacetime than Wheeler's first lawyer had let on. But these major breakthroughs are already several decades in the past, and one increasingly gets the impression that new impulses are needed. It may thus well be time to take another look at Wheeler's exploration of various degrees of lawfulness. It is hoped that the insights provided in this paper into the historical origins of Wheeler's fascinating reconfiguration of ideas can contribute to a better understanding of their position within the framework of science and their possible role in its future not least by highlighting the crucial questions that Wheeler ultimately struggled with: if no final theory of physics is to be had, on what ground can the lawfulness of the physical world stand? And how can physical law be reconciled with life, consciousness and even meaning on a \emph{global} scale?

\bibliography{habil}
\bibliographystyle{chicago}
\end{document}